\documentclass[aps,preprint,preprintnumbers,amsmath,amssymb]{revtex4}
\usepackage{amsmath,mathrsfs,amsbsy,graphicx,bm,amsthm,amsfonts}
\usepackage{units}
\usepackage{bbm}
\usepackage{times}
\usepackage{dcolumn}
\usepackage{mathrsfs}
\usepackage{amsmath,amssymb,epsfig}
\usepackage{color}
\usepackage{hyperref}
\hypersetup{colorlinks=true, 
	linkcolor=royalblue, 
	citecolor=magenta}

\newcommand\vertsp{\rule[-2mm]{1mm}{0mm} &}

\newcommand\morehorsp{\rule[-2.25mm]{0mm}{6mm}}

\def\aap{Astron. Astrophys.}
\def\mnras{Mon. Not. Roy. Astron. Soc.}
\def\apjl{Astrophys. J. Lett.}
\def\jcap{JCAP}

\newcommand{\LCDM}{\rm{\Lambda}CDM}

\hypersetup{
    colorlinks=true,
    linkcolor=red,
    citecolor=blue,
    urlcolor=magenta,
}
\begin{document}

\title{Multiple measurements on the cosmic curvature using  Gaussian process regression without calibration and a cosmological model}
\author{Xiaolong Gong$^{1}$, Yifei Xu$^{1}$, Tonghua Liu$^{1}$\footnote{Corresponding author: liutongh@yangtzeu.edu.cn},\\
Shuo Cao$^{2,\,\ 3}$\footnote{Corresponding author: caoshuo@bnu.edu.cn}, Jianyong Jiang$^{2}$, Yalong Nan$^{2,\,\ 3}$, Ruobin Ding$^{2,\,\ 3}$}
\affiliation{1. School of Physics and Optoelectronic Engineering, Yangtze University, Jingzhou 434023, China; \\
2. School of Physics and Astronomy, Beijing Normal University, Beijing 100875, China; \\ 3. Institute for Frontiers in Astronomy and Astrophysics, Beijing Normal University, Beijing 102206, China. }

\baselineskip=0.65 cm

\vspace*{-0.2cm}
\begin{abstract}
In this letter, we propose an improved cosmological model-independent method to measure cosmic curvature, combining the recent measurements of transverse and line-of-sight directions in the baryon acoustic oscillations (BAO) with cosmic chronometers (CC) datasets. Considering that the CC dataset is discrete and includes only 32 $H(z)$ measurements, we apply Gaussian process (GP) regression to fit the CC dataset and reconstruct them. Our methodology, which does not need the calibration or selection of any cosmological model, provide multiple measurements of spatial curvature ($\Omega_K$) at different redshifts (depending on the redshift coverage of BAO dataset). For combination of all BAO data, we find that the constraint result on cosmic curvature is $\Omega_K=-0.096^{+0.190}_{-0.195}$ with $1\sigma$ observational uncertainty.  Although the measured $\Omega_K$ is in good agreement with zero cosmic curvature within 1$\sigma$ confidence level, our result revels the fact of a closed universe. More importantly, our results show that the obtained $\Omega_K$ measurements are almost unaffected by different priors of the Hubble constant. This could help solve the issue of the Hubble tension that may be caused by inconsistencies in the spatial curvature  between the early and late universes.

\end{abstract}

\vspace*{0.2cm}

\maketitle
\section{Introduction}
The cosmic curvature, $\Omega_K$, is arguably the most fundamental and crucial parameter in our universe which affects the evolution of our universe and the dark energy properties. Any deviation from a flat universe would bring far-reaching thoughts and new understandings in the inflation theory and fundamental physics. Recently, some works suggested that spatial curvature of the universe is still an open issue \citep{2021PhRvD.103d1301H,2021APh...13102605D}.  More specifically,  the combination of {\it Planck} lensing and low redshift baryon acoustic oscillations (BAO) datasets revealed a flat universe with $\Omega_K=0.0007\pm0.0019$ \cite{2020A&A...641A...6P}. However, the work \citep{2020NatAs...4..196D} claimed that a close universe ($\Omega_K=-0.044^{+0.018}_{-0.015}$) was supported by the pure {\it Planck} Cosmic Microwave Background Radiation (CMBR) temperature and polarization power spectra data.  It should be emphasized that both methods depend, to some extent, on a particular cosmological model assumed, i.e. the non-flat $\Lambda$ cold dark matter ($\Lambda$CDM) model. This tension, like Hubble constant (more discussions on Hubble tension please see the references \cite{2019PhRvL.122v1301P,2019PhRvL.122f1105F,2021ApJ...912..150D,2022A&A...668A..51L,2020ApJ...895L..29L} ), their results are based on $\Lambda$CDM model.  Thus, cosmological model-independent curvature measurements can also test the $\Lambda$CDM model and are worth developing. Similarly, any curvature tension related to the CMB could be another evidence of $\Lambda$CDM model violation.

All these crises may be caused by measurement inconsistencies between the early universe and the late universe, and many works \citep{2023PhRvD.108f3522Q,2021ApJ...908...84V,2022JHEAp..33...10Z} point to the fact that relying on $\Lambda$CDM model will lead to such discrepancies.
To solve the crisis posed by the measurement of the curvature of the universe, a necessary approach is to confirm the value of $\Omega_K$ at late cosmic times using a method that is independence of the cosmological model. A non-exhaustive of references in this direction can be found in \citep{2019PhRvL.123w1101C,2020ApJ...897..127W,2021MNRAS.503.2179Q,2022ApJ...939...37L}. A typical curvature measurement idea is to obtain constraints on the geometrical quantity $\Omega_K$ using recent observational data provided by the high precision cosmological probes, namely, the type Ia supernovae (SNe Ia) distance modulus data, the cosmic chronometer and the radial BAO measurements of the Hubble parameter \citep{2022PhRvD.105f3516M}.  However, it should be noted that the BAO data are not completely cosmological model-independent, because one need to assume a prior on the radius of the sound horizon $r_s$, which is inferred from the CMB observations, i.e., the prior of $r_s$ obtained from the CMB analysis is used to calibrate the BAO distance. Secondly, previous work has generally used non-parametric Gaussian Processes (GP) method to reconstruct the expansion history of the universe. However, as mentioned in many works \cite{2018MNRAS.478.3640M,2022JCAP...12..029M}, one of the most recently debated topics for non-parametric reconstruction in cosmology with GP is that this technique is exposed to several foundational issues such as overfitting and kernel consistency problems \cite{2021arXiv210108565C}.

Inspired by the above, we propose an improved cosmological model-independent method to measure cosmic curvature combining the observations of BAO in the line-of-sight and transverse directions with CC datasets by using GP regression. The main advantage of this work is that it does not require the use of a priori values for the radius of the sound horizon $r_s$. We can perform measurements of cosmological model-independent curvature at multiple redshifts (depend on the BAO dataset). Moreover, the GP regression is powerful tool for reconstruction of function since the regression occurs in an infinite-dimensional function space without over-fitting problem \citep{Keeley0}. This paper is organized as follows: in Section 2 we briefly introduce the methodology and observational data used in our work. In Section 3, we show our results and discussion in Section 4.

\section{Methodology and observational data}
\subsection{Methodology for measuring on cosmic curvature}
The foundation of modern cosmology is based on the basic principles of cosmology, i.e., our universe is homogeneous and isotropic at large scales. The space-time of our universe can be described by Friedmann-Lematre-Robertson-Walker (FLRW) metric
\begin{equation}\label{eq1}
ds^2=dt^2-\frac{a(t)^2}{1-Kr^2}dr^2-a(t)^2r^2d\Omega^2,
\end{equation}
where $a(t)$ is the scale factor, and $K$ is dimensionless curvature taking one of three values $\{-1, 0, 1\}$ corresponding to a close, flat and open universe, respectively. The cosmic curvature parameter $\rm\Omega_K$ is related to $K$ and the Hubble constant $H_0$, as $\rm{\Omega_K}$ $=-c^2K/a^2_0H_0^2$.
The comoving distance $d_C$ is related to the evolution of the scale factor
\begin{equation}\label{eq2}
D_C=c\int^z_0\frac{dz'}{H(z')},
\end{equation}
where the redshift $z$ replaces the scale factor via relation $a=1/(1+z)$, the $c$ represents the speed of light, and the $H(z')$ denotes the Hubble parameter at redshift $z'$. However, we cannot actually obtain the comoving distance from observations. In fact, the cosmological distances we observe are luminosity or angular diameter distances. Since the luminosity distance is not used in this work, its relation to the comoving distance is not given. The comoving distance $D_C$ and the angular diameter distance $D_A$ satisfy the following relation \cite{Weinberg1972}
\begin{equation}\label{eq3}
{D_A}(z) = \left\lbrace \begin{array}{lll}
\frac{c}{(1+z)H_0\sqrt{|\rm\Omega_{\rm K}|}}\sinh\left[\sqrt{|\rm\Omega_{\rm K}|}{H_0D_C/c}\right]~~{\rm for}~~\rm\Omega_{K}>0,\\
D_C/(1+z)~~~~~~~~~~~~~~~~~~~~~~~~~~~~~~~~~~~~~~~~~~{\rm for}~~\rm\Omega_{K}=0, \\
\frac{c}{(1+z)H_0\sqrt{|\rm\Omega_{\rm K}|}}\sin\left[\sqrt{|\rm\Omega_{\rm K}|}H_0D_C/c\right]~~~~{\rm for}~~\rm\Omega_{K}<0.\\
\end{array} \right.
\end{equation}
As can be seen from the Eq. (\ref{eq3}), if we want to achieve the measurement of curvature in a cosmological model  independent way, one need to obtain the comoving distance and the angular diameter distance at the same redshift.
We will seek for cosmic chronometer (CC) and BAO observations to obtain the  comoving distances and  the angular diameter distances, respectively.

\subsection{Reconstructed comoving distance from cosmic chronometer observations}
Cosmic chronometers (CC) as the standard clocks provide the measurements of Hubble parameter, which derive from the combination of differential age estimates of systems with passively evolving star populations (e.g. globular clusters) and their corresponding spectral redshifts \cite{2002ApJ...573...37J},
\begin{equation}
H(z)=-\frac{1}{1+z}\frac{\Delta z}{\Delta t}.
\end{equation}
The CC data were obtained by measuring red envelope galaxies of different ages, i.e. different age method. The aging of stars can be seen as an indicator of the aging  of the universe. The spectra of stars can be converted into information about their age, because  the evolution of stars is well known. Since stars cannot be observed individually on a cosmic scale, it is common to use relatively homogeneous galaxy spectra of stellar populations. This approach relies on the detailed shape of the galaxy spectra rather than the luminosity of the galaxies. Therefore, CC data do not need to be calibrated and do not depend on underlying cosmological models \cite{2010JCAP...10..024A,2015PhRvD..92l3539L}.
The full information for total number of 32 CC is listed in Table 1 of the literature \cite{2002ApJ...573...37J}. The 32 CC data covers the redshift range of $0.07\leq z\leq1.965$ \cite{2014RAA....14.1221Z,2010JCAP...02..008S,2012JCAP...08..006M,2016JCAP...05..014M,2017MNRAS.467.3239R,2015MNRAS.450L..16M}.
The scatter diagram for 32 CC data is shown in the Fig.~1. At each redshift, the providing us with a continuous set of probability distribution functions (PDFs) at any given redshift.

\begin{figure}
\centering
\includegraphics[width=8.1cm,height=6cm]{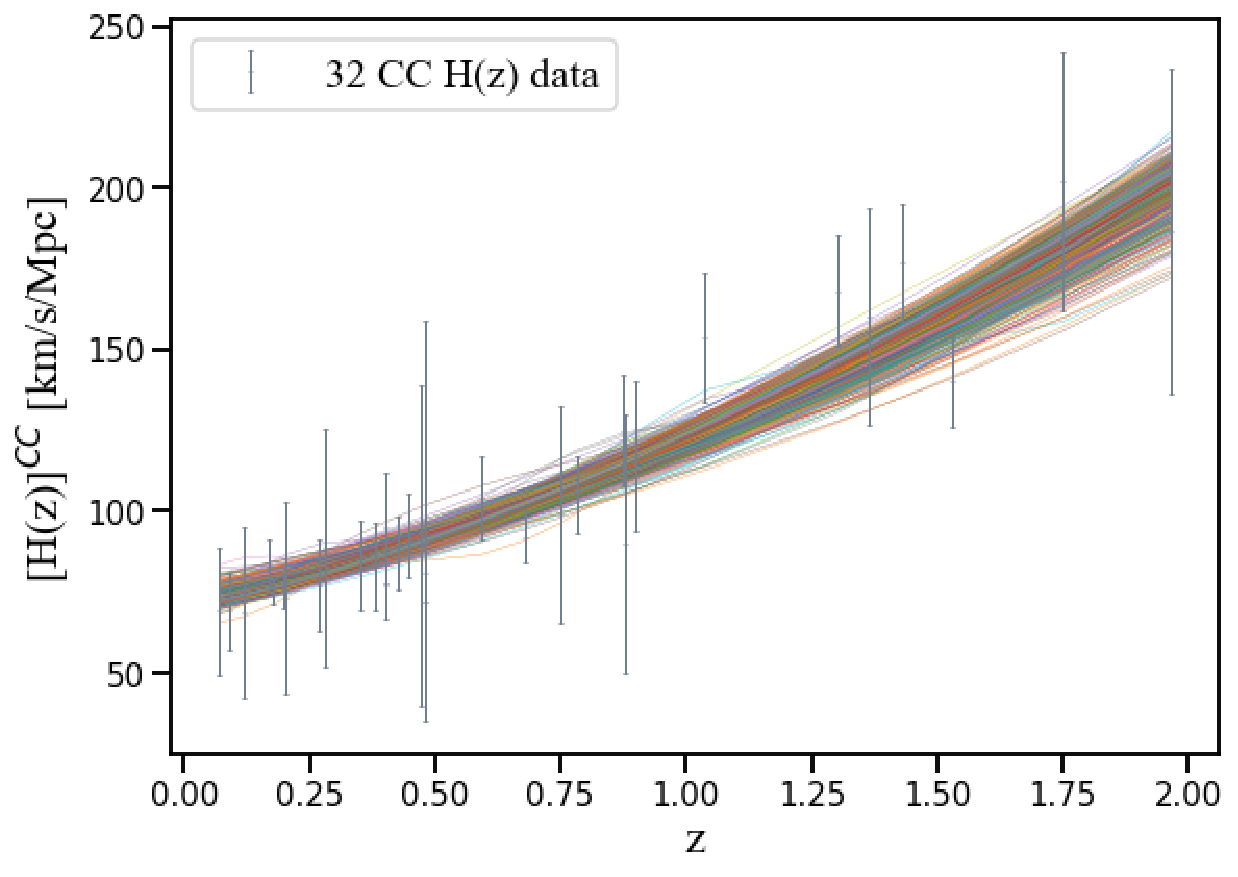}
\includegraphics[width=8.2cm,height=6cm]{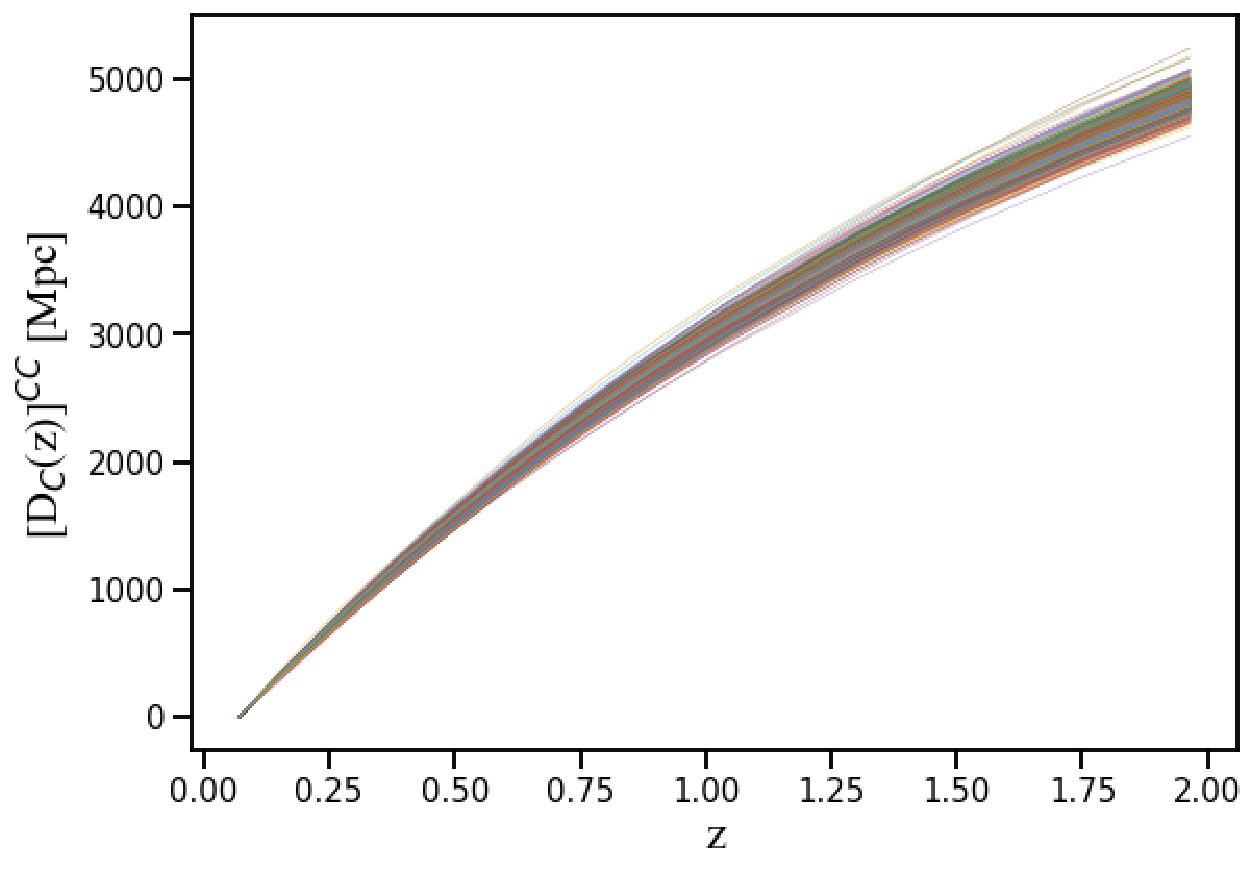}
\caption{\textit{Left panel}: The reconstructed 1000 realizations $H(z)$ from the CC dataset . \textit{Right panel}:  Based on reconstructed 1000 realizations of $[H(z)]^{\rm CC}$, the 1000 realizations  for comoving distances $[D_C(z)]^{\rm CC}$ obtained with the trapezoidal integral method.  }\label{fig:sn}
\end{figure}

To work with the redshift points corresponding to the BAO dataset, and we apply the Gaussian process regression to fit to CC dataset in order to reconstruct them at the chosen BAO redshifts.
For measuring curvature, we generate samples of the $H(z)$ from the posterior of the CC dataset. In order to perform the posterior sampling in a cosmological model-independent way, the GP regression~\citep{Holsclaw,Holsclaw1,Keeley0,ShafKimLind,ShafKimLind2,2019JCAP...12..035K} is considered here by using the code \texttt{GPHist}\footnote{https://github.com/dkirkby/gphist} \citep{GPHist}. The GP regression works by generating a large sample of functions $\gamma(z)$ determined by the covariance function.
The covariance between these functions can be described by a kernel function. We adopt a squared-exponential kernel to parameterize the covariance
\begin{equation}
    \langle \gamma(z_1)\gamma(z_2) \rangle = \sigma_f^2 \, \exp\{-[s(z_1)-s(z_2)]^2/(2\ell^2)\},
\end{equation}
with the hyperparameters $\sigma_f$ and $\ell$ that are marginalized over.  The $\gamma(z)$ is a random function inferred from the distribution defined by the covariance, and we adopt $\gamma(z) = \ln([H^{\rm fid}(z)/H_0]/[H(z)/H_0])$ to generate expansion histories $[H(z)]^{\rm CC}/H_0$ by using CC dataset.
Here $H^{\rm fid}(z)/H_0$ is chosen to be the best fit $\Lambda$CDM model for the CC dataset and serves the role of the mean function for GP regression.  Since the final reconstruction result is not completely independent of the mean function, it has some influence on the final reconstruction result because the value of the hyperparameter helps to track the deviation from the mean function. The true model should be very
close to flat $\LCDM$, hence, our choice for the mean function is very reasonable \citep{ShafKimLind,ShafKimLind2,Aghamousa2017}.  More detailed description of the hyperparameter,  kernel and and mean functions and the GP regression method can be found in the work \cite{2019ApJ...886L..23L,2023arXiv230913608L,2023arXiv230806951L,2021MNRAS.507..919L}.
It should be noted that we adopt a prior value on $H_0=74.03$ $km/s/Mpc$ from the \textit{Supernova $H_0$ for the Equation of State} (\textit{SH0ES}) collaboration \cite{2019ApJ...876...85R}. In fact, the a priori value of $H_0$ does not have much influence on the measurement of the curvature of the universe. We will discuss this point further later. The final reconstructed 1000 curves (or realizations) $[H(z)]^{\rm CC}$ from the CC dataset are shown in the left panel of the Fig.~\ref{fig:sn}.
By using a  trapezoidal integral method, we obtain the 1000 realizations for the comoving distances $[D_C(z)]^{\rm CC}$, and show in the right panel of the Fig.~\ref{fig:sn}.

\subsection{BAO observation}
For angular diameter distance $D_A$, we seek for seven BAO data points from the last data release of the BOSS and eBOSS collaboration \cite{2017MNRAS.470.2617A,2021MNRAS.500..736B,2020MNRAS.498.2492G,2020MNRAS.499.5527T,
2021MNRAS.501.5616D,2020MNRAS.499..210N,2021MNRAS.500.1201H,eBAO1,eBAO2}. The observations of BAO provide line-of-sight and transverse measurements, but these entangle with the radius of the sound horizon $r_s$, i.e., $D_M/r_s$ and $D_H/r_s$, where the $D_M=(1+z)D_A$ and $D_H=cH(z)^{-1}$  is transverse and line-of-sight directions measurement in BAO observations. To obtain the angular diameter distance, it is common procedure to use the prior value of radius of the sound horizon $r_s$ obtained from \textit{Planck} CMB observations. However, it is to some extent cosmological model-dependent, i.e., partly dependent on the assumptions of cosmological models. Therefore, we use the BAO observations in the transverse and line-of-sight directions, combined with observations from the $[H(z)]^{\rm CC}$ provided by the CC dataset, to obtain angular diameter distances independent of the cosmological model.
\begin{equation}\label{BAODA}
[D_A(z)]^{\rm BAO+CC}=\frac{c}{(1+z)\,[H(z)]^{\rm CC}}\cdot\frac{D_M}{r_s}\cdot\frac{r_s}{D_H}.
\end{equation}

For the BAO data, we do not perform any fit and use their seven redshift points as reference points.
Further, by combining  the observed data from BAO given in Table \ref{tab:BAOdata} with the GP of reconstructing 1000 $[H(z)]^{\rm CC}$ from the CC dataset, we obtain the five angular diameter distances at the chosen redshifts.
It should be noted that the redshift of CC dataset not full covers range of all BAO redshifts, considering the uncertainty of redshift extrapolation, we therefore rejected two BAO data point, i.e., $z=2.34$ and $z=2.35$. For the observational uncertainties of the BAO data, we assuming they follow gaussian PDFs (based on the symmetry of their 68\% confidence intervals), to obtain PDFs of $[D_A(z)]^{\rm BAO+CC}$ at the five redshifts.

\begin{table*}
	\centering
	\begin{tabular}{ccccc}
		\toprule
		Data source \vertsp $z$ \vertsp $D_M/r_s$ \vertsp $D_H/r_s$  \vertsp Reference  \\
		\hline\hline
		\morehorsp
		BOSS galaxy--galaxy \vertsp $ 0.38 $ \vertsp $ 10.27\pm 0.15 $ \vertsp $ 24.89\pm 0.58 $ \vertsp \cite{2017MNRAS.470.2617A} \\
		eBOSS galaxy--galaxy \vertsp $ 0.51 $ \vertsp $ 13.38\pm0.18 $ \vertsp $ 22.43 \pm 0.48 $ \vertsp \cite{2017MNRAS.470.2617A}\\
		\vertsp $ 0.70$ \vertsp $ 17.65 \pm 0.30 $ \vertsp $ 19.78\pm 0.46 $ \vertsp \cite{2021MNRAS.500..736B,2020MNRAS.498.2492G}\\
		\vertsp $ 0.85 $ \vertsp $ 19.50\pm 1.00 $ \vertsp $ 19.60\pm 2.10 $ \vertsp \cite{2020MNRAS.499.5527T,2021MNRAS.501.5616D}\\
		\vertsp $1.48$ \vertsp $ 30.21\pm0.79 $ \vertsp $ 13.23\pm 0.47 $ \vertsp \cite{2020MNRAS.499..210N,2021MNRAS.500.1201H}\\
		eBOSS Ly-$\alpha$--Ly-$\alpha$ \vertsp $ 2.34 $ \vertsp $ 37.41 \pm 1.86 $ \vertsp $ 8.86\pm0.29 $ \vertsp \cite{eBAO1} \\
		eBOSS Ly-$\alpha$--quasar \vertsp $ 2.35 $ \vertsp $ 36.30\pm1.80 $ \vertsp $ 8.20\pm0.36 $
\vertsp \cite{eBAO2} \\
		\hline\hline
	\end{tabular}
	\caption{The BAO data used in this work given in terms of $D_M = (1+z)D_A$ and $ D_H = c\,H(z)^{-1}$.}
	\label{tab:BAOdata}
\end{table*}

\section{Results and discussions}
The steps for constraining $\Omega_K$ are summarized as follows:

\begin{enumerate}
\item
    We draw 1000 reconstructed realizations $[H(z)]^{\rm CC}$ from CC dataset, and convert to comoving distances $[D_C(z)]^{\rm CC}$ by using the trapezoidal integral method at the five BAO redshifts based on Eq. (\ref{eq2}), respectively;

\item
    Based on Eq. (\ref{BAODA}), we calculate 1000 values of $[D_A(z)]^{\rm BAO+CC}$ at the five BAO redshifts by combing the 1000 $[H(z)]^{\rm CC}$ realizations and the five BAO observations, respectively;
\item
   We combine the 1000 realizations of the $[D_C(z)]^{\rm CC}$ with the realizations of $[D_A(z)]^{\rm BAO+CC}$ to get the estimations of the PDFs of $\Omega_K$ at the five BAO redshifts based on the Eq. (\ref{eq3}), respectively;
\item
    With the above procedure, the uncertainty on $[D_C(z)]^{\rm CC}$ and $[D_A(z)]^{\rm BAO+CC}$  , as well as their mutual correlation are intrinsically included in the respective PDFs. Then, we use these
    samples to reconstruct the multi-variate distribution of the five $\Omega_K$ measurements.

\item Finally, we extract the marginal distribution of the individual  $\Omega_K$  thus leaving us with five uncorrelated PDFs.
\end{enumerate}

\begin{figure}
\centering
{\includegraphics[width=8cm,height=6.3cm]{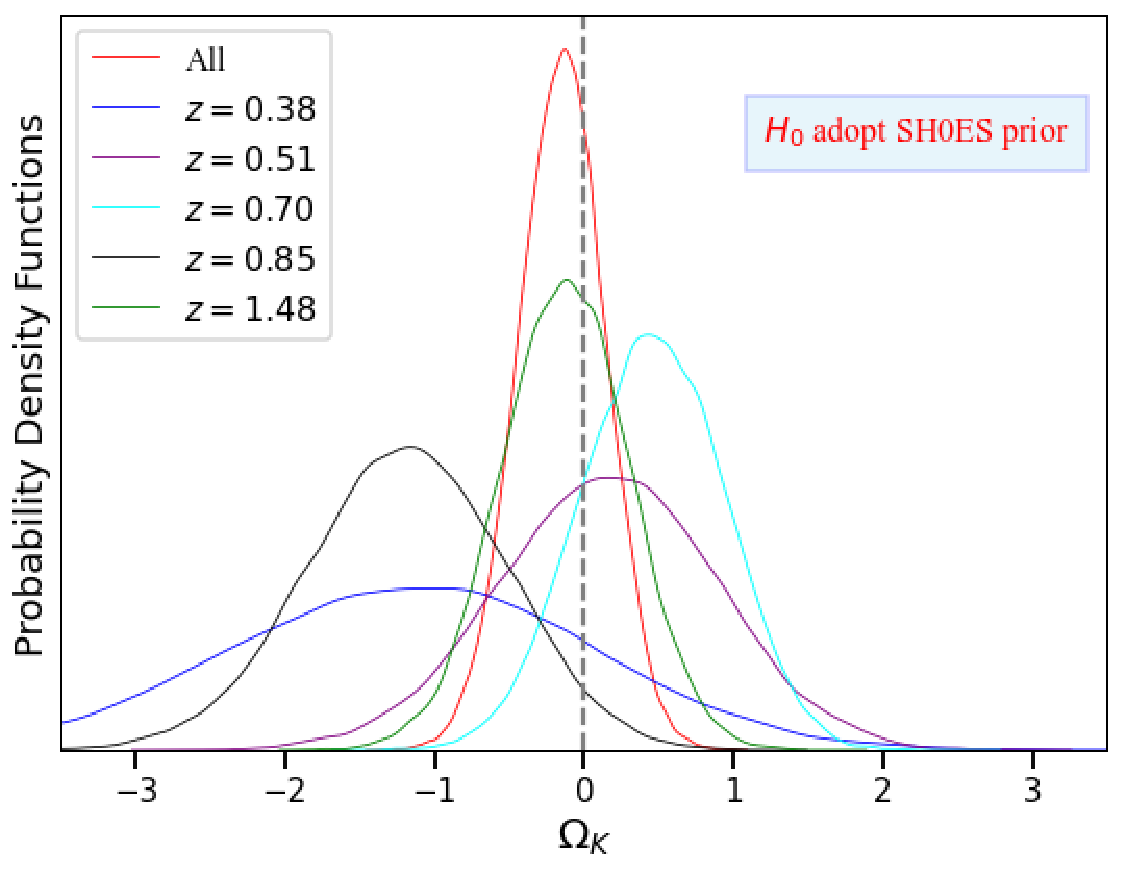}
\includegraphics[width=8cm,height=6.3cm]{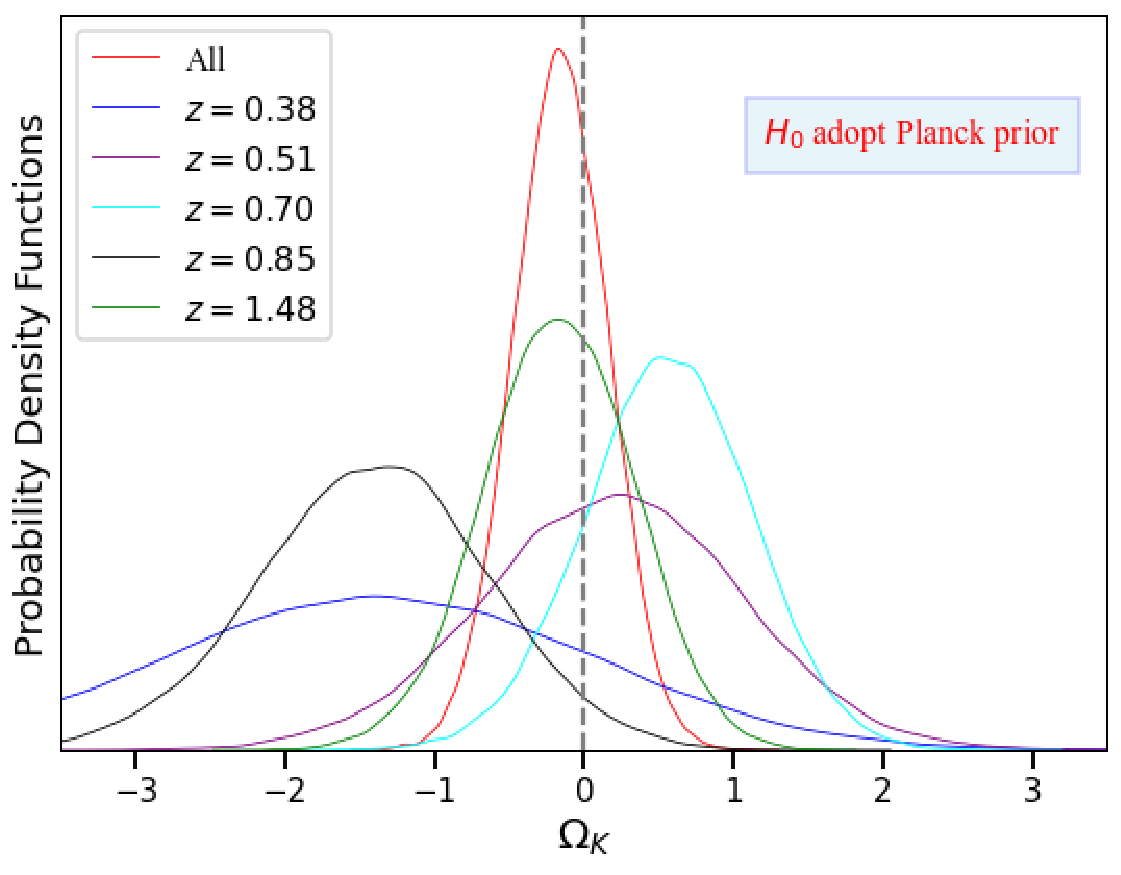}}
\caption{\textit{Left panel}: The individual measurement on $\Omega_K$ using the $H_0=74.03$ $km/s/Mpc$ prior reported by the \textit{SH0ES} collaboration. \textit{Right panel}: The individual measurement on $\Omega_K$ using the $H_0=67.4$ $km/s/Mpc$  prior reported by the \textit{Planck} CMB observations.}\label{fig3}
\end{figure}

The final step is similar to the process used in Markov Chain Monte Carlo (MCMC) to derive constraints on the likelihood parameter and to allow for a direct conversion of the correlation and uncertainty of the PDFs of  the distances into the PDFs of $\Omega_K$. The whole method is similar to a combination of MCMC-like parameter estimation and Gaussian process reconstruction, which is called as Gaussian Process Monte Carlo
(GPMC). The GPMC method was first proposed by the work \cite{2023PhRvD.107b3520R}.

In order to perform the above procedure, we have to take a prior value for the Hubble constant. We first adopt a prior value on $H_0=74.03$ $km/s/Mpc$ reported by the \textit{SH0ES} collaboration. The individual $\Omega_K$ PDFs and the final results obtained combining them are reported in the left panel of the Fig. \ref{fig3}. We note that our results for $\Omega_K$ are individual measurements, which means that we achieve measurements of $\Omega_K$ at different redshifts. Although the final result obtained combining five individual $\Omega_K$ PDFs reports that our universe has a flat spatial curvature with the $\Omega_K=-0.096^{+0.190}_{-0.195}$ at $1\sigma$ confidence level, the precision of our constrained results on $\Omega_K$ is not particularly high.
It is reasonable to get such precision that we used only five points of BAO data.
Furthermore, we obtain a negative value for the centre of curvature, which is consistent with recently reported work, i.e., combining the \textit{Planck} temperature and polarization power spectra data, the work showed that a closed universe ($\rm\Omega_K=-0.044^{+0.018}_{-0.015}$) was supported. Our work also reinforces that the universe is likely to be a closed universe.

Some recent works suggested that the $H_0$ tension could possibly be caused by the inconsistency of spatial curvature between the early-universe and late-universe \cite{2020NatAs...4..196D,2021PhRvD.103d1301H,2021APh...13102605D}.  An effective way to study this problem is to simultaneously constrain the curvature of the universe and the Hubble constant, and to use a cosmological model-independent method \citep{2019PhRvL.123w1101C,2020ApJ...897..127W,2021MNRAS.503.2179Q,2022ApJ...939...37L}.
This is due to the strong degeneracy between the cosmic curvature and the Hubble constant. Inspired by these, we further adopt a prior value on $H_0=67.4$ $km/s/Mpc$ reported by the \textit{Planck} CMB observations. The final results of the PDFs of $\Omega_K$ are shown in the right panel of the Fig. \ref{fig3}. The result is $\Omega_K=-0.101^{+0.207}_{-0.212}$ at $1\sigma$ confidence level by combining five individual $\Omega_K$ PDFs. Compared with the results using the \textit{SH0ES} prior, our results show that the obtained values of curvature are almost unaffected with the different prior of Hubble constant by using our method.
This conclusion can also be seen from the Eq. \ref{eq2}, the combination of $[D_A(z)]^{BAO+CC}$ requires the $[H(z)]^{CC}$ component. If we write $H(z)$ in the form of the $H_0*E(z)$, where the $E(z)$ is dimensionless Hubble parameter, we can see that the Hubble constant on the left and right sides of the equation is cancelled out.
Irrespective of the prior of $H_0$ used, our results demonstrate the validity and robustness of the method of measuring $\Omega_K$ in this work.

As a final remark, we also look forward to a large amount of future data, not only from the observations of BAO, but also from the cosmic chronometer, allowing us to further improve the precision of $\Omega_K$ measurements.  Future surveys, such as Dark Energy Spectroscopic Instrument (DESI), will measure $H(z)r_s$ at several redshifts and be able to bring down those errors, thus to improve the constrain on the cosmic curvature in a wider range and with much higher precision and accuracy without cosmological model assumptions by using our method. Meanwhile, considering the existence of many data reconstruction methods, such as machine learning methods, the rapid development and technical improvement of these methods makes us optimistic about measuring curvature with higher  precision and accuracy in the future.

\section{Acknowledgments}

Liu. T.-H was supported by National Natural Science Foundation of China under Grant No. 12203009; Chutian Scholars Program in Hubei Province (X2023007).  Cao S. was supported by the National Natural Science Foundation of China under Grants No. 12021003, 11633001, and 11920101003; the Strategic Priority Research Program of the Chinese Academy of Sciences, Grant No. XDB23000000; and the Interdiscipline Research Funds of Beijing Normal University.

\bibliographystyle{unsrt}

\end{document}